\documentclass[aps,pra,twocolumn,superscriptaddress,floatfix,nofootinbib]{revtex4-2}

\usepackage[T1]{fontenc}
\usepackage[utf8]{inputenc}
\usepackage{amsmath,amssymb,amsfonts,mathtools,bm}
\usepackage{graphicx}
\usepackage{xcolor}
\usepackage{booktabs}
\usepackage{orcidlink}
\usepackage{hyperref}
\hypersetup{
  colorlinks=true,
  linkcolor=blue!55!black,
  citecolor=blue!55!black,
  urlcolor=blue!55!black
}
\allowdisplaybreaks

\newcommand{\ii}{\mathrm{i}}
\newcommand{\dd}{\mathrm{d}}
\newcommand{\ee}{\mathrm{e}}
\newcommand{\OmP}{\Omega_P}
\newcommand{\OmS}{\Omega_S}
\newcommand{\Oma}{\Omega_a}
\newcommand{\D}{\Delta}
\newcommand{\G}{\Gamma}
\newcommand{\Psc}{P_{\mathrm{sc}}}
\newcommand{\order}{\mathcal{O}}
\newcommand{\half}{\tfrac{1}{2}}
\newcommand{\ket}[1]{|#1\rangle}
\newcommand{\bra}[1]{\langle #1|}
\newcommand{\braket}[2]{\langle #1|#2\rangle}
\newcommand{\ketbra}[2]{|#1\rangle\langle #2|}

\begin{document}

\title{Bright-state source cancellation in dissipative shortcut Raman atom optics}

\author{Asad Ali~\!\!\orcidlink{0000-0001-9243-417X}}
\email{asal68826@hbku.edu.qa}
\affiliation{Qatar Center for Quantum Computing, College of Science and Engineering, Hamad Bin Khalifa University, Doha, Qatar}

\author{Saif Al-Kuwari~\!\!\orcidlink{0000-0002-4402-7710}}
\email{smalkuwari@hbku.edu.qa}
\affiliation{Qatar Center for Quantum Computing, College of Science and Engineering, Hamad Bin Khalifa University, Doha, Qatar}

\author{M. I. Hussain~\!\!\orcidlink{0000-0002-6231-7746}}
\affiliation{Qatar Center for Quantum Computing, College of Science and Engineering, Hamad Bin Khalifa University, Doha, Qatar}

\author{H. Kuniyil~\!\!\orcidlink{0000-0003-0338-1278}}
\affiliation{Qatar Center for Quantum Computing, College of Science and Engineering, Hamad Bin Khalifa University, Doha, Qatar}

\author{M.~T.~Rahim\orcidlink{0000-0003-1529-928X}} 
\affiliation{Qatar Center for Quantum Computing, College of Science and Engineering, Hamad Bin Khalifa University, Doha, Qatar}

\author{Saeed Haddadi~\!\!\orcidlink{0000-0002-1596-0763}}
\email{haddadi@ipm.ir}
\affiliation{School of Particles and Accelerators, Institute for Research in Fundamental Sciences (IPM), P.O. Box 19395-5531, Tehran, Iran}

\begin{abstract}
Spontaneous Raman scattering limits shortcut-assisted atom optics, but its microscopic origin is obscured once the lossy excited state is adiabatically eliminated. We organize the problem around a single quantity: in the instantaneous dark--bright basis the lower-manifold optical source is carried entirely by the bright-state amplitude, $S=\Omega b$, so that primary spontaneous scattering reduces to the compact functional $P_{\mathrm{sc}}=2\!\int\!\kappa|b|^2\,dt$. This recovers the known dissipative-stimulated Raman adiabatic passage (STIRAP) loss in transparent form and makes the action of a shortcut explicit: ideal counterdiabatic stimulated Raman shortcut-to-adiabatic passage (STIRSAP) cancels the bright-state \emph{source}, not the optical decay coefficient. We show this cancellation is exact in the full three-level model at the counterdiabatic point $\Omega_a=2\dot\theta$, for arbitrary one-photon detuning, Rabi frequency, and pulse duration. The residual source splits into orthogonal quadratures---shortcut mismatch (real) and two-photon Doppler detuning (imaginary)---which invites a velocity-selective protocol that nulls the Doppler quadrature for a chosen momentum class with a second, phase-shifted lower-state field. Our central result is that this source nulling is never superior to simply chirping the two-photon detuning: the two coincide only when the selected class $\delta_c$ is small compared with the bright-state gap $|\mu|=\Omega^2/4\Delta$, and the nulling degrades and then fails as $\delta_c\to|\mu|$---precisely the regime of launched or warm clouds and high-order large-momentum-transfer (LMT) optics that motivates velocity selection. The controlling quantity is the magnitude of the residual Hamiltonian perturbation a scheme leaves behind, not the residual source it cancels. As a complement to existing multi-pulse decay budgets, we cast a single-pulse mode-error budget for LMT interferometry entirely in terms of the bright-state source, and delineate when shortcut-assisted Raman control reduces the total scattering cost.
\end{abstract}

\maketitle

\section{Introduction}
\label{sec:intro}

Stimulated Raman adiabatic passage (STIRAP) is one of the most powerful tools in coherent quantum control because it transfers population through an instantaneous dark state while strongly suppressing occupation of a lossy intermediate level~\cite{Bergmann1998,vitanov2001,vitanov2017}. In a three-level $\Lambda$ system the transfer is governed by the geometry of the dark-state path rather than the accumulated two-photon pulse area, making it resilient to moderate pulse-area imperfections~\cite{oreg1984}. That robustness is especially valuable in atom optics, where coherent Raman processes manipulate internal and momentum states simultaneously. Large-momentum-transfer (LMT) interferometry is a key setting: enlarging the momentum separation between arms increases the enclosed spacetime area and thereby the inertial sensitivity~\cite{mcguirk2000,muller2008,chiow2011}. The same optical fields that drive coherent momentum transfer also produce spontaneous emission, ac Stark shifts, Doppler-dependent errors, and pulse-efficiency loss, all of which accumulate over many pulses and limit contrast~\cite{cronin2009,butts2013,kotru2014,kotru2015}.

The central dissipative question is not whether the dark-state path is followed, but what microscopic quantity controls spontaneous Raman scattering once the excited state is eliminated. Open-system treatments of spontaneous emission are well established through Lindblad, quantum-jump, and no-jump methods~\cite{dalibard1992,molmer1993}, and dissipative STIRAP has been analyzed by master-equation and adiabatic-elimination approaches~\cite{ivanov2005,reiter2012}. In parallel, shortcut-to-adiabaticity techniques --- counterdiabatic driving and shortcut STIRAP --- reproduce exact dark-state following in finite time in closed systems~\cite{demirplak2003,berry2009,torrontegui2013,chen2010,li2016shortcut,guery2019}, have been realized with cold atoms~\cite{du2016}, and have been studied under dissipation and dephasing~\cite{blekos2020,shi2021effect,funo2021general}. A spontaneous-decay error budget for multi-pulse Raman LMT has also been formulated via the Lindblad equation~\cite{chrostoski2024}, and counterdiabatic Raman atom optics with the shortcut encoded directly into the pulse envelopes has been analyzed for compact gravimetry~\cite{Ali2026companion}. What has remained less transparent is a compact, source-level quantity that controls primary Raman scattering and permits a direct comparison between bare STIRAP, ideal direct stimulated Raman shortcut-to-adiabatic passage (STIRSAP), and optically reconstructed shortcuts in one dissipative framework. In particular, where Ref.~\cite{shi2021effect} studies how spontaneous emission affects shortcut-to-adiabatic passage in three-level systems, here we identify the bright-state source $S=\Omega b$ that controls the loss, show the counterdiabatic cancellation to be \emph{exact} in the full three-level model for arbitrary $\D$, $\Omega$, $\G$, and $T_p$ (Appendix~\ref{app:exact}), and use the same quadrature decomposition to settle the velocity-class addressing question.

Here we provide such a description. The organizing result, obtained after transformation to the instantaneous dark--bright basis, is that the lower-manifold optical source feeding the primary scattering channel is carried entirely by the bright-state amplitude,
\begin{equation}
  S(t)=\OmP(t)c_g(t)+\OmS(t)c_a(t)=\Omega(t)b(t),
  \label{eq:intro-Sbright}
\end{equation}
so that the reduced primary-scattering functional collapses to
\begin{equation}
  P_{\mathrm{sc}}^{\mathrm{red}}
  =2\int_0^{T_p}\kappa(t)\,|b(t)|^2\,\dd t .
  \label{eq:intro-Pred}
\end{equation}
The dark state is lossless; the bright component carries the primary scattering channel. This makes the action of shortcut driving transparent: in bare STIRAP, finite-time rotation of the dark state generates a nonzero bright-state source and hence residual scattering; in ideal direct STIRSAP, the counterdiabatic coupling cancels that \emph{source}, not the optical decay coefficient. We verify the cancellation directly in the full three-level model with a concrete counterdiabatic field.

The same formalism separates two physically distinct error mechanisms. Shortcut mismatch appears as a \emph{real} bright-state source quadrature, while two-photon Doppler detuning appears as an \emph{imaginary} quadrature. This separation invites a velocity-selective protocol in which a phase-controlled lower-state shortcut nulls the Doppler quadrature for a chosen momentum class $\delta_c$. We analyze this construction carefully and reach a definite, and partly cautionary, conclusion: source nulling is never better than the standard technique of chirping the two-photon detuning to bring the selected class onto resonance~\cite{kotru2015}. The two coincide exactly when $\delta_c$ is small compared with the bright-state gap $|\mu|=\Omega^2/4\Delta$; as $\delta_c$ grows toward $|\mu|$ the nulling degrades and then fails, while the chirp stays robust. The reason is structural --- both leave the same residual \emph{source}, but the chirp leaves a Hamiltonian perturbation of order the detuning offset, whereas the nulling leaves a fixed perturbation of order $\delta_c$ that is fine-tuned to cancel only for the matched class. Robustness is therefore governed by the perturbation magnitude, not by the residual source. We regard this boundary, together with the source identity, Eq.~\eqref{eq:intro-Sbright}, and the exact counterdiabatic cancellation, as the main contribution of the present work.

Finally, we use the framework to formulate a single-pulse mode-error budget for LMT optics that combines coherent transfer error, residual primary scattering, auxiliary reconstruction loss, and Doppler curvature, and we state explicitly when shortcut-assisted Raman control reduces the total scattering cost. This part is a design-level extension of the single-pulse theory rather than a complete many-pulse interferometer treatment, and it overlaps in scope with Refs.~\cite{chrostoski2024,Ali2026companion}; we delineate what is distinct.

The paper is organized so that the main text states physical results and defers algebra to the appendices. Section~\ref{sec:model} sets up the dissipative model and the exact primary-scattering functional. Section~\ref{sec:bright} establishes the bright-state source identity, the reduced loss functional, and its numerical validation for bare STIRAP. Section~\ref{sec:cd} gives the exact counterdiabatic source cancellation and its full-model confirmation. Section~\ref{sec:doppler} presents the velocity-class nulling-versus-chirp analysis. Section~\ref{sec:budget} develops the implementation cost and LMT budget. Sections~\ref{sec:discussion} and \ref{sec:conclusion} discuss and conclude. Appendices~\ref{app:normloss}--\ref{app:exact} contain the derivations.

\begin{figure}[t]
    \centering
    \includegraphics[width=\linewidth]{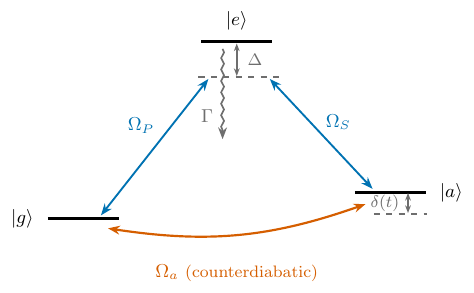}
\caption{Three-level $\Lambda$ system: long-lived states $\ket{g}$ and $\ket{a}$
coupled to a lossy excited state $\ket{e}$ (linewidth $\Gamma$) by pump
($\Omega_P$) and Stokes ($\Omega_S$) fields at one-photon detuning $\Delta$
(dashed line: virtual level) and time-dependent two-photon detuning $\delta(t)$.
The counterdiabatic field $\Omega_a$ drives $\ket{g}\!\leftrightarrow\!\ket{a}$
directly.}
\label{fig:placeholder}
\end{figure}

\section{Dissipative model and the primary-scattering functional}
\label{sec:model}

We use a minimal three-level Raman $\Lambda$ model that isolates the primary dissipative channel associated with the optically excited intermediate state (see Fig.~\ref{fig:placeholder}). The long-lived lower states $\ket{g}$ and $\ket{a}$ (which in LMT optics may also carry momentum labels differing by two photon recoils) are coupled to a decaying excited state $\ket{e}$ by a pump field $\OmP(t)$ and a Stokes field $\OmS(t)$. A direct lower-state shortcut coupling $\Oma(t)$ and a residual two-photon detuning $\delta(t)$ complete the model; $\delta(t)$ collects the uncompensated Doppler shift of the addressed momentum class together with any residual differential light shift or chirp error. In the rotating-wave approximation, with $\OmP,\OmS,\Oma$ real envelopes, the Hamiltonian in the ordered basis $\{\ket{g},\ket{e},\ket{a}\}$ is
\begin{equation}
  H_3(t)=\hbar
  \begin{pmatrix}
    0 & \OmP/2 & +\ii\Oma/2 \\
    \OmP/2 & -\D & \OmS/2 \\
    -\ii\Oma/2 & \OmS/2 & \delta
  \end{pmatrix},
  \label{eq:H3}
\end{equation}
where $\D$ is the one-photon detuning. The lower-state coupling is written with imaginary off-diagonal elements so that a real $\Oma$ is Hermitian; equivalently
\begin{equation}
  H_a=\ii\hbar\frac{\Oma}{2}\bigl(\ketbra{g}{a}-\ketbra{a}{g}\bigr).
  \label{eq:Ha}
\end{equation}
This sign convention is fixed so that, after transformation to the instantaneous dark--bright basis (Sec.~\ref{sec:bright}), the direct shortcut condition takes the form $\Oma(t)=2\dot\theta(t)$; the same convention is used consistently in the reduced equations and in the numerical benchmark of Appendix~\ref{app:numerics}.

Spontaneous emission from $\ket{e}$ is described by jump operators $L_\mu=\sqrt{\Gamma_\mu}\,\ket{\ell_\mu}\bra{e}$ with $\sum_\mu\Gamma_\mu=\G$, and the density operator obeys the Lindblad equation
\begin{equation}
  \dot\rho
  =-\frac{\ii}{\hbar}[H_3(t),\rho]
  +\sum_\mu\!\left(L_\mu\rho L_\mu^\dagger-\half\{L_\mu^\dagger L_\mu,\rho\}\right).
  \label{eq:lindblad}
\end{equation}
Because $\sum_\mu L_\mu^\dagger L_\mu=\G\ketbra{e}{e}$, the detailed channel structure enters the no-jump branch only through the total linewidth $\G$; post-jump recycling is not modeled, and any emission event is later treated as removal from the desired coherent mode. The no-jump (non-Hermitian) evolution then follows from $H_{\mathrm{nh}}=H_3-\tfrac{\ii\hbar\G}{2}\ketbra{e}{e}$, and the unnormalized amplitudes $c_{g,e,a}$ obey
\begin{align}
  \dot c_g &= -\ii\tfrac{\OmP}{2}c_e+\tfrac{\Oma}{2}c_a,
  \label{eq:cg}\\
  \dot c_e &= -\ii\tfrac{\OmP}{2}c_g+(\ii\D-\tfrac{\G}{2})c_e-\ii\tfrac{\OmS}{2}c_a,
  \label{eq:ce}\\
  \dot c_a &= -\tfrac{\Oma}{2}c_g-\ii\tfrac{\OmS}{2}c_e-\ii\delta\,c_a .
  \label{eq:ca}
\end{align}
These are exact within the assumed three-level, no-jump model and serve as the starting point for all reductions.

The no-jump norm $N(t)=|c_g|^2+|c_e|^2+|c_a|^2$ is the survival weight of the coherent branch, and $1-N(t)$ is the probability that at least one spontaneous-emission event has occurred. A short calculation (Appendix~\ref{app:normloss}) gives the exact norm-loss identity
\begin{equation}
  \dot N(t)=-\G|c_e(t)|^2,
  \qquad
  P_{\mathrm{sc}}=\G\int_0^{T_p}|c_e(t)|^2\,\dd t .
  \label{eq:Psc-exact}
\end{equation}
Equation~\eqref{eq:Psc-exact} is the operational definition of primary spontaneous scattering used throughout. To connect it to Raman physics, we eliminate the excited state. The exact memory-kernel solution of Eq.~\eqref{eq:ce} and its short-memory (local Raman) reduction are derived in Appendix~\ref{app:local}; the result is the instantaneous relation
\begin{equation}
  c_e(t)\simeq\frac{S(t)}{2\zeta},
  \quad
  S(t)=\OmP c_g+\OmS c_a,
  \quad
  \zeta=\D+\tfrac{\ii}{2}\G,
  \label{eq:ce-local}
\end{equation}
valid under the scale-separation conditions $T_p|\zeta|\gg1$, $|\dot\Omega|/(\Omega|\zeta|)\ll1$, and $|\dot\theta|/|\zeta|\ll1$, together with $|S|/(2|\zeta|)\ll1$. Equation~\eqref{eq:ce-local} shows that the excited state follows the lower-manifold optical source $S(t)$ almost instantaneously, and that primary scattering is controlled by $|S|^2$.

\section{Bright-state source and the reduced loss functional}
\label{sec:bright}

\subsection{Source identity and compact functional}

Resolving the lower manifold into the instantaneous dark and bright directions selected by the Raman fields is the key step. With the total Rabi scale $\Omega=\sqrt{\OmP^2+\OmS^2}$ and the mixing angle $\theta$ defined by $\sin\theta=\OmP/\Omega$, $\cos\theta=\OmS/\Omega$, the dark and bright states are
\begin{equation}
  \ket{D}=\cos\theta\,\ket{g}-\sin\theta\,\ket{a},
  \quad
  \ket{B}=\sin\theta\,\ket{g}+\cos\theta\,\ket{a}.
  \label{eq:DB}
\end{equation}
Writing $\ket{\psi_g}=d\ket{D}+b\ket{B}$ and substituting into $S=\OmP c_g+\OmS c_a$ (Appendix~\ref{app:source}) yields the central identity
\begin{equation}
  S(t)=\Omega(t)\,b(t).
  \label{eq:Sbright}
\end{equation}
The dark amplitude cancels from the source because $\ket{D}$ has no optical coupling to $\ket{e}$; the entire source is carried by the bright amplitude. Combining Eqs.~\eqref{eq:Psc-exact}, \eqref{eq:ce-local}, and \eqref{eq:Sbright} gives the reduced loss functional in compact form,
\begin{equation}
  P_{\mathrm{sc}}^{\mathrm{red}}
  =2\int_0^{T_p}\kappa(t)\,|b(t)|^2\,\dd t,
  \label{eq:Psc-kappa}
\end{equation}
with the bright-state decay and light-shift coefficients and the complex response rate
\begin{equation}
  \kappa=\frac{\Omega^2\G}{8(\D^2+\G^2/4)},
  \quad
  \Lambda=\frac{\Omega^2\D}{4(\D^2+\G^2/4)},
  \quad
  \mu=\kappa+\ii\Lambda .
  \label{eq:kappaLambdaMu}
\end{equation}
Within the reduced local Raman model, primary scattering is thus governed by the bright-state population rather than the dark-state population. The far-detuned bright-state gap $|\mu|\simeq\Lambda\simeq\Omega^2/4\D$ will set the scale against which all source-cancellation schemes below must be measured.

\subsection{Bare STIRAP and numerical validation}

\begin{figure}[t]
\centering
\includegraphics[width=\columnwidth]{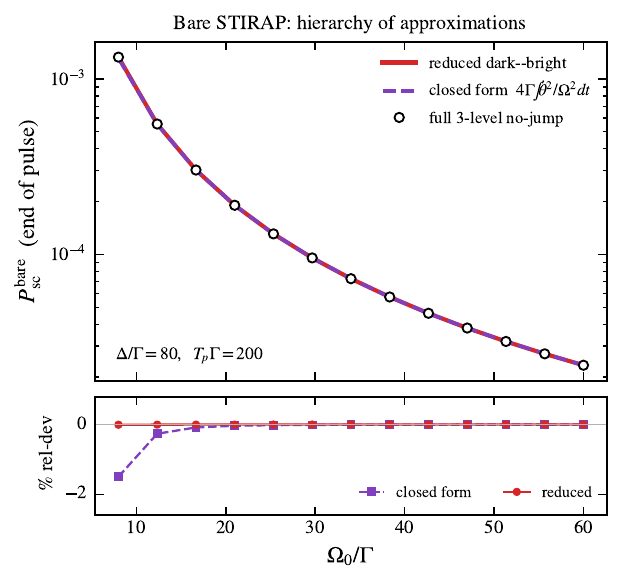}
\caption{Bare-STIRAP scattering across the hierarchy of approximations: full three-level no-jump dynamics (open circles), reduced dark--bright propagation (solid), and the closed-form leading estimate Eq.~\eqref{eq:Pbare} (dashed), versus the optical Rabi scale $\Omega_0/\G$. The lower panel shows the relative deviation of the reduced and closed-form results from the full model. Parameters: $\D/\G=80$, $T_p\G=200$, and Gaussian $\dot\theta$ schedule with $\sigma=T_p/6$ (see Appendix~\ref{app:gaussian}).}
\label{fig:bare-hierarchy}
\end{figure}

For bare STIRAP ($\Oma=0$, $\delta=0$), the only mechanism that creates bright amplitude is the finite-time rotation of the basis. The exact moving-basis equations are derived in Appendix~\ref{app:bare}; the bright amplitude obeys $\dot b=\dot\theta\,d-\mu b$, and in the local-response limit $b\simeq\dot\theta/\mu$. Using the identity $2\kappa/|\mu|^2=4\G/\Omega^2$ (Appendix~\ref{app:bare}), the leading bare-STIRAP scattering estimate is
\begin{equation}
  P_{\mathrm{sc}}^{\mathrm{bare}}
  \simeq
  4\G\int_0^{T_p}\frac{\dot\theta^2(t)}{\Omega^2(t)}\,\dd t .
  \label{eq:Pbare}
\end{equation}
Figure~\ref{fig:bare-hierarchy} validates the full chain of approximations. The full three-level no-jump dynamics [Eqs.~\eqref{eq:cg}--\eqref{eq:ca} with Eq.~\eqref{eq:Psc-exact}], the reduced dark--bright propagation [$\dot b=\dot\theta d-\mu b$ with $\dot P_{\mathrm{sc}}=2\kappa|b|^2$], and the closed form~\eqref{eq:Pbare} agree across the optical Rabi scale. The reduced model tracks the full model to better than $10^{-4}$ in relative terms, and the closed form to about $1.5\%$. This confirms the source identity and the local Raman elimination in the regime of Eq.~\eqref{eq:ce-local}.

\section{Counterdiabatic cancellation of the primary source}
\label{sec:cd}

\subsection{Exact cancellation condition}

For a real dark-state path, the counterdiabatic Hamiltonian is $H_{\mathrm{cd}}=\ii\hbar\dot\theta(\ketbra{D}{B}-\ketbra{B}{D})$, realized in the lower manifold by the coupling $\Oma$ of Eq.~\eqref{eq:Ha}. Collecting the moving-basis term, the Raman bright-state response, the shortcut coupling, and the residual detuning (Appendix~\ref{app:source-full}) gives the exact reduced bright-state equation
\begin{equation}
  \dot b+\bigl[\mu(t)+\ii\delta(t)\cos^2\theta(t)\bigr]b=F(t)\,d,
  \label{eq:b-source}
\end{equation}
with the residual dark-to-bright source
\begin{equation}
  F(t)=\dot\theta(t)-\frac{\Oma(t)}{2}+\ii\,\delta(t)\sin\theta(t)\cos\theta(t).
  \label{eq:Fsource}
\end{equation}
For real $\dot\theta$, $\Oma$, and $\delta$, the source separates into orthogonal quadratures $F=F_R+\ii F_I$ with $F_R=\dot\theta-\Oma/2$ and $F_I=\delta\sin\theta\cos\theta$, so that shortcut mismatch and Doppler detuning do not interfere at leading order:
\begin{equation}
  |F|^2=\Bigl(\dot\theta-\tfrac{\Oma}{2}\Bigr)^2+\delta^2\sin^2\theta\cos^2\theta .
  \label{eq:Fsq}
\end{equation}
On two-photon resonance ($\delta=0$), the counterdiabatic condition
\begin{equation}
  \Oma(t)=2\dot\theta(t)
  \label{eq:CDcond}
\end{equation}
gives $F=0$. Equation~\eqref{eq:b-source} then reduces to $\dot b+\mu b=0$, and for an initially dark state ($b(0)=0$) the unique solution is $b(t)=0$, so the primary Raman channel through the original excited state is exactly suppressed,
\begin{equation}
  P_{\mathrm{sc}}^{\mathrm{primary}}=0 ,
  \label{eq:Pprimary-zero}
\end{equation}
within the reduced local Raman model. This is a dynamical \emph{source}-cancellation effect: the optical decay coefficient $\kappa$ is unchanged; what vanishes is the bright-state population it multiplies. In the local-response regime, the general residual source yields
\begin{equation}
  P_{\mathrm{sc}}^{\mathrm{primary}}
  \simeq
  4\G\int_0^{T_p}\frac{|F(t)|^2}{\Omega^2(t)}\,\dd t ,
  \label{eq:Pprimary}
\end{equation}
which we use below for both shortcut mismatch and residual detuning. The residual source $F$ retains exactly this form in the full three-level model, without any adiabatic elimination, as we show by constructing the exact dark--bright--excited generator in Appendix~\ref{app:exact}.

\subsection{Full-model confirmation}

\begin{figure*}[t]
\centering
\includegraphics[width=\textwidth]{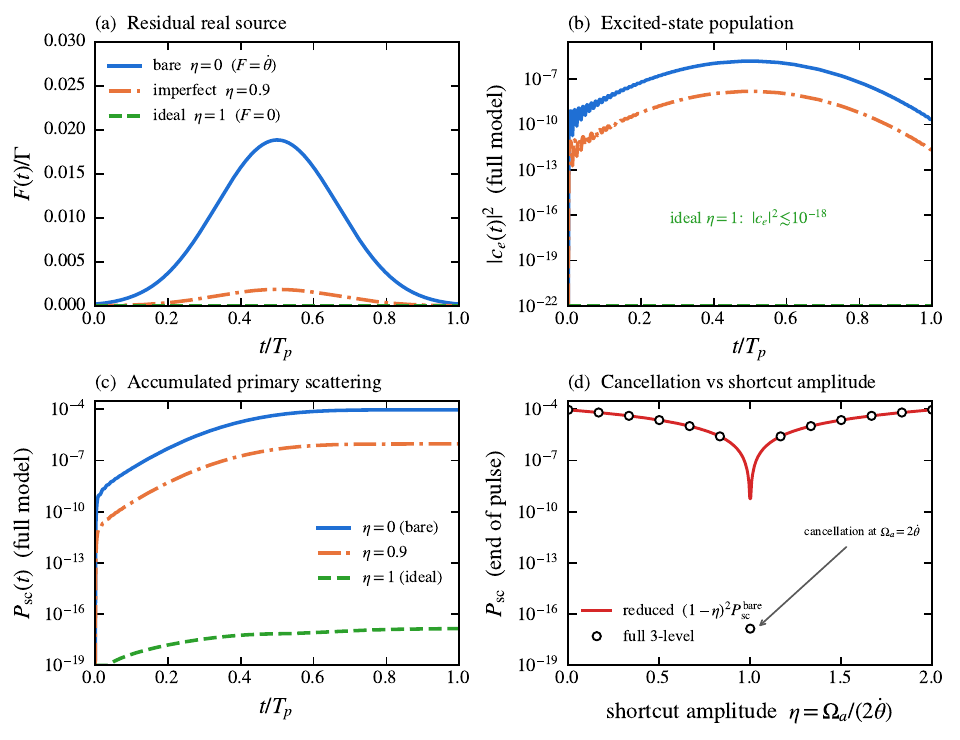}
\caption{Full-model confirmation of bright-state source cancellation, with the counterdiabatic field parametrized as $\Oma=\eta\,2\dot\theta$. (a) Residual real source $F=(1-\eta)\dot\theta$ for bare STIRAP ($\eta=0$), imperfect STIRSAP ($\eta=0.9$), and ideal STIRSAP ($\eta=1$). (b) Full-model excited-state population $|c_e(t)|^2$; for $\eta=1$ it remains below $10^{-18}$. (c) Accumulated primary scattering $P_{\mathrm{sc}}(t)$ from the full no-jump model. (d) End-of-pulse scattering versus shortcut amplitude $\eta=\Oma/(2\dot\theta)$: the full three-level result (markers) lies on the reduced prediction $(1-\eta)^2P_{\mathrm{sc}}^{\mathrm{bare}}$ (line) and collapses to $\sim10^{-17}$ at the counterdiabatic point $\eta=1$. Parameters as in Fig.~\ref{fig:bare-hierarchy}.}
\label{fig:source-cancellation}
\end{figure*}

Equation~\eqref{eq:Pprimary-zero} is a statement about the \emph{reduced} model; an essential check is that a concrete counterdiabatic field cancels scattering in the \emph{full} three-level dynamics, with the correct sign convention. Figure~\ref{fig:source-cancellation} confirms this. We parametrize the shortcut amplitude as $\Oma=\eta\,2\dot\theta$ and integrate the full no-jump equations [Eqs.~\eqref{eq:cg}--\eqref{eq:ca}]. Panel (a) shows the residual real source $F=(1-\eta)\dot\theta$ for $\eta=0$ (bare), $\eta=0.9$ (imperfect), and $\eta=1$ (ideal). Panels (b) and (c) show that the full-model excited-state population and the accumulated primary scattering collapse toward machine precision as $\eta\to1$. Panel (d) is the decisive test: the full-model end-of-pulse scattering follows the reduced parabola $(1-\eta)^2P_{\mathrm{sc}}^{\mathrm{bare}}$ across the whole range and reaches $\sim10^{-17}$ at the counterdiabatic point $\eta=1$. The quadratic dependence on the mismatch $1-\eta$ confirms that residual scattering is controlled by $|F|^2$ exactly as in Eq.~\eqref{eq:Pprimary}, and the collapse at $\eta=1$ confirms the exact cancellation (see Eq.~\eqref{eq:Pprimary-zero}) in the full model.

\section{Velocity-class source nulling versus frequency chirping}
\label{sec:doppler}

\subsection{Two routes to a selected velocity class}

After the real source is cancelled by Eq.~\eqref{eq:CDcond}, a residual two-photon detuning $\delta$ leaves the imaginary Doppler source
\begin{equation}
  F_D(t)=\ii\,\delta\,\sin\theta(t)\cos\theta(t),
  \label{eq:FD}
\end{equation}
which reopens the primary channel even when the nonadiabatic source is exactly removed. From Eq.~\eqref{eq:Pprimary},
\begin{equation}
  P_{\mathrm{sc}}^{D}\simeq 4\G\,\delta^2 J_D,
  \qquad
  J_D=\int_0^{T_p}\frac{\sin^2\theta\cos^2\theta}{\Omega^2}\,\dd t,
  \label{eq:PscD}
\end{equation}
a parabola in $\delta$ centered at $\delta=0$. To make a selected momentum class $\delta_c$ lossless, there are two routes as follows:

\emph{(i) Frequency chirp.}\ The standard technique in Raman atom optics is to ramp the two-photon detuning so the selected class sits on resonance~\cite{kotru2015}, i.e.\ an atom of true detuning $\delta$ experiences $\delta_{\mathrm{eff}}=\delta-\delta_c$ while the single-quadrature counterdiabatic coupling Eq.~\eqref{eq:CDcond} is retained. The selected atom ($\delta_{\mathrm{eff}}=0$) transfers without primary scattering; a neighbor $\delta=\delta_c+\varepsilon$ has source $\ii\varepsilon\sin\theta\cos\theta$ and response rate $\mu+\ii\varepsilon\cos^2\theta\simeq\mu$, so $P_{\mathrm{sc}}\simeq4\G\varepsilon^2J_D$. The key feature is that the Hamiltonian differs from the resonant dark-state Hamiltonian only by a perturbation of order $\varepsilon$, small for every nearby class.

\emph{(ii) Two-quadrature source nulling.}\ Alternatively one keeps the lasers at fixed frequency and adds a second, phase-shifted lower-state quadrature. Replacing the single real shortcut by the general dark--bright coupling
\begin{equation}
  H_q
  =\hbar\frac{\Omega_x}{2}\bigl(\ketbra{D}{B}+\ketbra{B}{D}\bigr)
  +\ii\hbar\frac{\Omega_y}{2}\bigl(\ketbra{D}{B}-\ketbra{B}{D}\bigr),
  \label{eq:Hq}
\end{equation}
the residual source becomes (Appendix~\ref{app:nulling})
\begin{equation}
  F_q(t)=\dot\theta-\frac{\Omega_y}{2}
  +\ii\Bigl[\delta\sin\theta\cos\theta-\frac{\Omega_x}{2}\Bigr].
  \label{eq:Fq}
\end{equation}
Choosing $\Omega_y=2\dot\theta$ and $\Omega_x=2\delta_c\sin\theta\cos\theta$ cancels both quadratures for the matched class $\delta=\delta_c$, giving $F_q=0$ and $b(t)=0$. For a neighbor $\delta=\delta_c+\varepsilon$ the residual source is again purely imaginary, $F_q=\ii\varepsilon\sin\theta\cos\theta$, identical to the chirp. At leading order $b\simeq F_q/\mu$, both routes therefore predict the same shifted parabola
\begin{equation}
  P_{\mathrm{sc}}^{\mathrm{null}}\simeq4\G\,(\delta-\delta_c)^2 J_D .
  \label{eq:Pnull}
\end{equation}

\subsection{The nulling offers no advantage and has a sharp ceiling}

Equation~\eqref{eq:Pnull} suggests that source nulling reproduces the velocity selectivity of a chirp. It does --- but only within the local-response approximation $b\simeq F/\mu$, i.e.\ for $\delta_c\ll|\mu|$, and even there it requires amplitude \emph{and} phase control of an additional lower-state field [the quadrature shortcut of Eq.~\eqref{eq:Hq}, with $\Omega_x$ carrying both an off-diagonal $g$--$a$ coupling and diagonal energy modulations $\pm\delta_c\sin^2 2\theta/2$; see Appendix~\ref{app:nulling}], whereas a chirp uses a single, ubiquitous frequency knob. The nulling thus achieves, at best, parity with a simpler control.

Beyond the small-detuning regime, the two routes diverge sharply, and the nulling fails. The reason is structural. Both leave the \emph{same} residual source $\propto\varepsilon$, but they leave very different Hamiltonian perturbations. The chirp leaves an $\order(\varepsilon)$ perturbation; the nulling leaves a \emph{fixed} $\order(\delta_c)$ perturbation --- the diagonal modulation of Eq.~\eqref{eq:Hq} plus the full physical detuning --- that is fine-tuned to cancel only for the matched class. Nothing protects that cancellation once $\delta_c$ becomes comparable to the bright-state gap $|\mu|$ that enforces adiabatic dark-state following; for any mismatch $\varepsilon$ the bright population then grows beyond the local-response estimate and the transfer itself breaks. Robustness is controlled by the perturbation magnitude, not by the residual source.

\begin{figure*}[t]
\centering
\includegraphics[width=\textwidth]{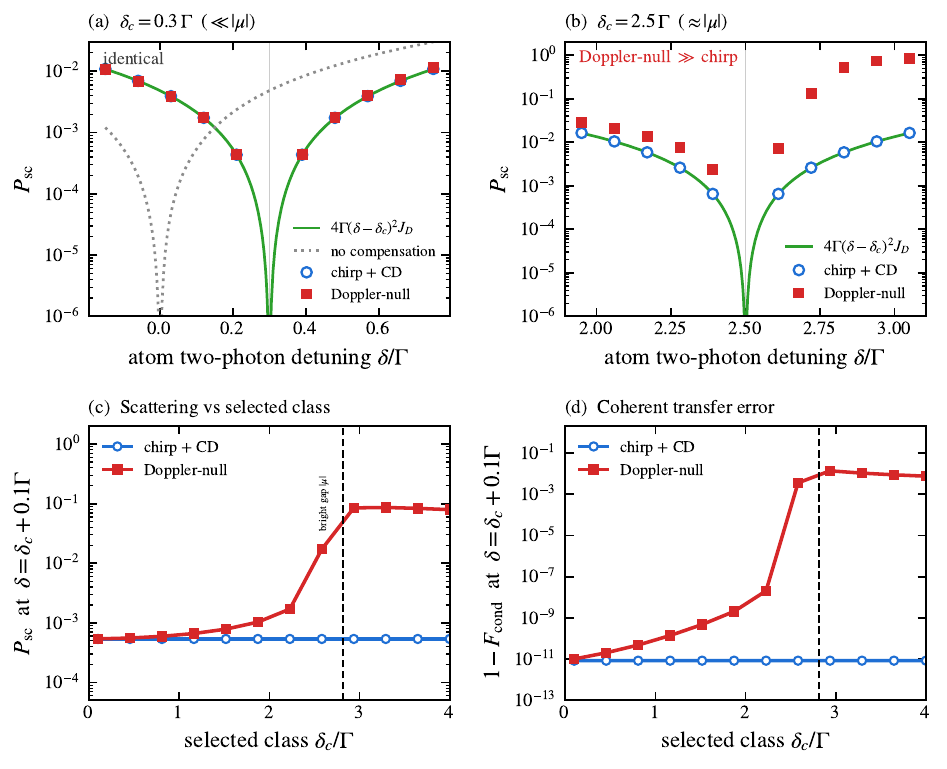}
\caption{Velocity-class source nulling versus frequency chirping, in the full three-level model ($|\mu|=\Omega_0^2/4\D=2.81\,\G$). (a) Primary scattering versus atom detuning $\delta$ for a small selected class $\delta_c=0.3\,\G\ll|\mu|$: chirp (open circles) and nulling (filled squares) coincide and follow the shifted parabola Eq.~\eqref{eq:Pnull} (line); the dotted curve is the uncompensated case, centered at $\delta=0$. (b) For $\delta_c=2.5\,\G\approx|\mu|$, the nulling scatters $10$--$100\times$ more than the chirp for off-resonant atoms. (c) Scattering at a fixed offset $\delta=\delta_c+0.1\,\G$ versus the selected class $\delta_c$: the chirp is flat, while the nulling diverges as $\delta_c$ crosses the bright-state gap $|\mu|$ (dashed line). (d) The corresponding coherent transfer error $1-F_{\mathrm{cond}}$: the chirp stays near $10^{-11}$, while the nulling's coherent transfer collapses beyond $|\mu|$. Parameters as in Fig.~\ref{fig:bare-hierarchy}.}
\label{fig:doppler-chirp}
\end{figure*}

Figure~\ref{fig:doppler-chirp} demonstrates this directly in the full three-level model, with $|\mu|=\Omega_0^2/4\D=2.81\,\G$ for these parameters. (a) For $\delta_c=0.3\,\G\ll|\mu|$, the chirp and nulling results coincide on the parabola Eq.~\eqref{eq:Pnull}: equivalent, as anticipated. (b) For $\delta_c=2.5\,\G\approx|\mu|$, the chirp still tracks the parabola while the nulling scatters one to two orders of magnitude more for off-resonant atoms. (c) Holding a fixed small offset $\varepsilon=0.1\,\G$ and scanning the selected class, the chirp scattering is essentially independent of $\delta_c$, whereas the nulling tracks it only for $\delta_c\lesssim|\mu|/2$ and then rises by more than two orders of magnitude as $\delta_c$ crosses $|\mu|$. (d) The coherent transfer error confirms the mechanism: the chirp maintains $1-F_{\mathrm{cond}}\sim10^{-11}$ across the whole range, while the nulling loses coherent transfer entirely once $\delta_c>|\mu|$.

The conclusion is unambiguous: two-quadrature source nulling is never superior to chirping the two-photon detuning. It is exactly equivalent where the local-response approximation holds ($\delta_c\ll|\mu|$), and progressively --- then catastrophically --- worse as the addressed class approaches the bright-state gap. Because velocity selectivity matters most precisely when the addressed Doppler shifts are appreciable (launched or warm clouds, and high-order LMT where the class spacing grows), the nulling fails in the regime that motivates it, while the chirp remains robust. The shifted-parabola statement given in Eq.~\eqref{eq:Pnull} is therefore valid only in a narrow neighborhood and only for $\delta_c\ll|\mu|$. The orthogonality of the two source quadratures remains a useful organizing statement --- a real counterdiabatic term cannot cancel a detuning-induced coupling, and vice versa --- but it does not translate into a practical advantage over standard frequency control.

\section{Implementation cost and LMT mode budget}
\label{sec:budget}

\subsection{Optical reconstruction cost}

The ideal direct shortcut $\Oma$ of Eq.~\eqref{eq:Ha} is treated as a Hermitian lower-state interaction with no intrinsic scattering cost. Whether such a coupling is physically available is implementation-dependent and, for LMT, restrictive: since $\ket{g}$ and $\ket{a}$ differ in momentum by two photon recoils, the counterdiabatic $g\!\leftrightarrow\!a$ coupling must itself impart $2\hbar k$ and therefore cannot be a momentum-free microwave; it must be realized optically. When the shortcut is synthesized by a separate auxiliary Raman pair through an additional excited state (linewidth $\Gamma_R$, one-photon detuning $\Delta_R$), it carries its own scattering channel. For a balanced dark-path reconstruction (Appendix~\ref{app:aux}),
\begin{equation}
  P_{\mathrm{aux}}=\frac{\pi\Gamma_R\Delta_R}{2(\Delta_R^2+\Gamma_R^2/4)}
  \xrightarrow{\ \Delta_R\gg\Gamma_R\ }\frac{\pi\Gamma_R}{2\Delta_R}.
  \label{eq:Paux}
\end{equation}
An important caveat constrains how this term is counted. If, instead of a separate auxiliary pair, the counterdiabatic correction is absorbed into reshaped pump and Stokes envelopes~\cite{Ali2026companion}, the reshaped fields scatter through the \emph{same} excited state with the \emph{same} coefficient $\kappa$, and there is no separate $P_{\mathrm{aux}}$ channel: the cost is already inside the primary functional Eq.~\eqref{eq:Psc-kappa}. Equation~\eqref{eq:Paux} therefore applies specifically to a separate-auxiliary-pair realization; the envelope-encoded realization is accounted for entirely by the residual primary source $F(t)$.

\subsection{Single-pulse mode error and contrast}

For LMT optics the relevant figure of merit is the survival of a specified coherent output mode after many momentum-transfer operations. Writing the conditional coherent transfer error $\epsilon_{\mathrm{coh}}=1-F_{\mathrm{cond}}$ with $F_{\mathrm{cond}}=|c_a(T_p)|^2/N(T_p)$, and combining it additively with the scattering channels and the Doppler curvature (Appendix~\ref{app:budget}), the single-pulse mode error is
\begin{equation}
  \epsilon_{\mathrm{mode}}
  =\epsilon_{\mathrm{coh}}^{(0)}
  +P_{\mathrm{primary}}^{\mathrm{resid},0}
  +P_{\mathrm{aux}}
  +A_D^{\mathrm{mode}}\sigma_D^2 ,
  \label{eq:mode-budget}
\end{equation}
where the superscript $0$ denotes the designed central class, $\sigma_D$ is the Doppler width, and the small-detuning mode-error curvature separates into coherent-endpoint and scattering parts, $A_D^{\mathrm{mode}}=M_D+4\G J_D$ (Appendix~\ref{app:budget}). For $N_\pi$ statistically independent pulses the contrast follows $C/C_0\simeq\exp[-N_\pi\epsilon_{\mathrm{mode}}]$, and a target contrast imposes the design inequality $\epsilon_{\mathrm{mode}}\le-N_\pi^{-1}\ln C_{\mathrm{target}}$. This budget is a design-level small-error estimate; it does not include interferometer phase noise, wavefront aberration, beam-splitter imbalance, or many-pulse coherent correlations, and it overlaps in scope with Refs.~\cite{chrostoski2024,Ali2026companion}. What is distinct here is that each scattering term is expressed through the bright-state source: $P_{\mathrm{primary}}^{\mathrm{resid}}$ via $|F|^2$, the Doppler curvature via $J_D$, and the implementation cost via Eq.~\eqref{eq:Paux} or, for envelope encoding, via $F$ itself.

A shortcut protocol improves the LMT budget only when the reduction in primary source-induced scattering and coherent transfer error exceeds the additional auxiliary and Doppler-related costs. For ideal direct cancellation the residual primary term vanishes and the comparison reduces to suppressed bare scattering versus auxiliary (or envelope-encoding) cost; the break-even condition is
\begin{equation}
  4\G\!\int_0^{T_p}\!\frac{\dot\theta^2}{\Omega^2}\,\dd t
  \approx
  P_{\mathrm{aux}}
  +4\G\!\int_0^{T_p}\!\frac{|F_{\mathrm{resid}}|^2}{\Omega^2}\,\dd t .
  \label{eq:breakeven}
\end{equation}

\section{Discussion}
\label{sec:discussion}

The organizing result of this work is the separation between the bright-state decay coefficient and the bright-state source. The coefficient $\kappa(t)$ is fixed by the Raman intensity, detuning, and excited-state linewidth, and is present whenever the fields are applied. The source $F(t)$ depends on the protocol and determines whether the bright state is populated at all. Bare STIRAP has a nonzero source because the dark state rotates in finite time; ideal direct STIRSAP cancels that source within the reduced local Raman model. The reduction in primary scattering from bare STIRAP to ideal STIRSAP is therefore a dynamical source-cancellation effect, not a reduction of the optical decay coefficient --- a distinction we confirmed directly in the full model (Fig.~\ref{fig:source-cancellation}).

The same source decomposition cleanly separates shortcut mismatch (a real quadrature) from two-photon Doppler detuning (an imaginary quadrature), and this is where our analysis is most cautionary. The decomposition invites a two-quadrature protocol that nulls the Doppler source for a selected class, and within the leading local-response approximation that protocol reproduces a shifted scattering parabola. But a controlled comparison shows it never beats the standard practice of chirping the two-photon detuning: the two coincide for $\delta_c\ll|\mu|$ and the nulling fails as $\delta_c\to|\mu|$ (Fig.~\ref{fig:doppler-chirp}). The lesson generalizes beyond this system. When a control cancels a dissipative source by fine-tuned interference, its robustness is set by the size of the residual Hamiltonian perturbation it leaves behind, not by the residual source it cancels; a scheme that leaves an $\order(\delta_c)$ perturbation to cancel an $\order(\delta_c)$ source is fragile, whereas a resonance condition that leaves an $\order(\varepsilon)$ perturbation is robust. We therefore present the quadrature decomposition as a diagnostic organizing principle and the chirp as the practical method, rather than advancing source nulling as a protocol improvement.

For implementation, the momentum selection rule for LMT means the direct counterdiabatic coupling is not a free resource: it must be optical and therefore carries either the auxiliary cost Eq.~\eqref{eq:Paux} or, if encoded in reshaped Raman envelopes~\cite{Ali2026companion}, a cost already contained in the primary source $F$. The mode-error budget Eq.~\eqref{eq:mode-budget} and break-even condition Eq.~\eqref{eq:breakeven} make the resulting trade-offs explicit and connect to existing multi-pulse error analyses~\cite{chrostoski2024}. The reduced loss functional and the design-level contrast extrapolation hold within the local Raman model; the cancellation of the primary channel itself, however, is exact in the full three-level model at the counterdiabatic point (Appendix~\ref{app:exact}), independent of that reduction. In practice, approximate or envelope-encoded realizations of the shortcut leave a small residual infidelity even at the counterdiabatic point, consistent with the concrete single-pulse fidelities reported for envelope-encoded shortcuts~\cite{Ali2026companion}.

\section{Conclusion}
\label{sec:conclusion}

We have developed a dissipative bright-state framework for Raman STIRAP and shortcut-assisted atom optics. From the Lindblad equation for a lossy three-level $\Lambda$ system, we obtained the exact no-jump norm-loss identity, the controlled local Raman reduction, and the dark--bright source identity $S=\Omega b$, which collapses the primary-scattering problem to the compact functional $P_{\mathrm{sc}}=2\!\int\!\kappa|b|^2\,dt$. Within this description, primary spontaneous scattering is governed by the bright-state amplitude; bare STIRAP scatters as $4\G\!\int\!\dot\theta^2/\Omega^2\,dt$, and ideal direct STIRSAP cancels the bright-state source at $\Oma=2\dot\theta$, which we confirmed --- with the correct sign convention --- directly in the full three-level model.

Using the orthogonal-quadrature structure of the source, we analyzed velocity-class addressing and found that two-quadrature Doppler source nulling is never superior to chirping the two-photon detuning: the two are identical for $\delta_c\ll|\mu|$ and the nulling degrades and then fails as the selected class approaches the bright-state gap $|\mu|=\Omega^2/4\D$, because robustness is set by the residual Hamiltonian perturbation rather than the residual source. We closed with a single-pulse mode-error budget and break-even condition for LMT optics that express each error channel through the bright-state source, and that delineate when shortcut-assisted Raman control reduces the total scattering cost and when auxiliary or Doppler-related effects dominate. Beyond this setting, the bright-state source perspective --- isolating the amplitude that carries loss as the first step toward suppressing it, while tracking the perturbation that controls robustness --- may be useful in other dissipative shortcut-driven control problems.

\section*{Acknowledgments}
This publication was partially supported by the Qatar Research, Development and Innovation (QRDI) Council under the Academic Research Grant ARG01-0603-230468. The findings and views expressed herein are solely the responsibility of the authors.

\appendix

\section{No-jump norm-loss identity}
\label{app:normloss}

The no-jump state $\ket{\psi_{\mathrm{nj}}}=c_g\ket{g}+c_e\ket{e}+c_a\ket{a}$ evolves under $\ii\hbar\partial_t\ket{\psi_{\mathrm{nj}}}=H_{\mathrm{nh}}\ket{\psi_{\mathrm{nj}}}$ with $H_{\mathrm{nh}}=H_3-\tfrac{\ii\hbar\G}{2}\ketbra{e}{e}$. Differentiating $N=\braket{\psi_{\mathrm{nj}}}{\psi_{\mathrm{nj}}}$, we have
\begin{equation}
  \dot N=\frac{\ii}{\hbar}\bra{\psi_{\mathrm{nj}}}\bigl(H_{\mathrm{nh}}^\dagger-H_{\mathrm{nh}}\bigr)\ket{\psi_{\mathrm{nj}}}.
\end{equation}
Since $H_3$ is Hermitian, $H_{\mathrm{nh}}^\dagger-H_{\mathrm{nh}}=\ii\hbar\G\ketbra{e}{e}$, and using $\braket{e}{\psi_{\mathrm{nj}}}=c_e$, we obtain
\begin{equation}
  \dot N(t)=-\G|c_e(t)|^2,
\end{equation}
which integrates to Eq.~\eqref{eq:Psc-exact}.

\section{Excited-state memory kernel and local Raman approximation}
\label{app:local}

With $\zeta=\D+\tfrac{\ii}{2}\G$ and $S=\OmP c_g+\OmS c_a$, Eq.~\eqref{eq:ce} reads $\ii\dot c_e=\tfrac{S}{2}-\zeta c_e$. The integrating factor $\ee^{-\ii\zeta t}$ gives the exact solution
\begin{equation}
  c_e(t)=\ee^{\ii\zeta t}c_e(0)-\frac{\ii}{2}\int_0^t \ee^{\ii\zeta(t-s)}S(s)\,\dd s,
  \label{eq:ce-memory}
\end{equation}
with kernel $\ee^{\ii\zeta(t-s)}=\ee^{\ii\D(t-s)}\ee^{-\G(t-s)/2}$: oscillatory memory at the one-photon scale, damped on the timescale $\G^{-1}$. For $c_e(0)=0$ and substitution $u=t-s$,
\begin{equation}
  c_e(t)=-\frac{\ii}{2}\int_0^t \ee^{\ii\zeta u}S(t-u)\,\dd u .
\end{equation}
When $S$ varies slowly on the memory timescale, $S(t-u)=S(t)-u\dot S(t)+\cdots$, and with $I_n=\int_0^\infty u^n\ee^{\ii\zeta u}\,\dd u$, $I_0=\ii/\zeta$, the leading term is $c_e\simeq S/(2\zeta)$ [Eq.~\eqref{eq:ce-local}], the first correction being $\order(\dot S/\zeta^2)$. Sufficient conditions are $T_p|\zeta|\gg1$, $|\dot\Omega|/(\Omega|\zeta|)\ll1$, $|\dot\theta|/|\zeta|\ll1$, and $|S|/(2|\zeta|)\ll1$.

\section{Dark--bright reduction and the source identity}
\label{app:source}

Substituting $c_e\simeq S/(2\zeta)$ into Eqs.~\eqref{eq:cg} and \eqref{eq:ca} closes the lower-manifold dynamics with
\begin{equation}
  H_{\mathrm{red}}=\hbar
  \begin{pmatrix}
    \OmP^2/(4\zeta) & \OmP\OmS/(4\zeta)+\ii\Oma/2 \\[0.4em]
    \OmP\OmS/(4\zeta)-\ii\Oma/2 & \delta+\OmS^2/(4\zeta)
  \end{pmatrix}.
  \label{eq:Hred}
\end{equation}
Using $1/\zeta=\D/(\D^2+\G^2/4)-\ii(\G/2)/(\D^2+\G^2/4)$, the anti-Hermitian part is the rank-one loss operator $-\ii\hbar\,\tfrac{\G}{8(\D^2+\G^2/4)}\,\bigl(\begin{smallmatrix}\OmP^2&\OmP\OmS\\ \OmP\OmS&\OmS^2\end{smallmatrix}\bigr)$, while $\Oma$ stays Hermitian and does not enter the loss. The norm decay gives $P_{\mathrm{sc}}^{\mathrm{red}}=\tfrac{\G}{4(\D^2+\G^2/4)}\int_0^{T_p}|S|^2\,\dd t$ with $S=\OmP c_g+\OmS c_a$. With $\ket{\psi_g}=d\ket{D}+b\ket{B}$ and Eq.~\eqref{eq:DB}, $c_g=d\cos\theta+b\sin\theta$ and $c_a=-d\sin\theta+b\cos\theta$, so
\begin{equation}
  S=d(\OmP\cos\theta-\OmS\sin\theta)+b(\OmP\sin\theta+\OmS\cos\theta)=\Omega b,
\end{equation}
which is Eq.~\eqref{eq:Sbright}. With $\kappa$ as in Eq.~\eqref{eq:kappaLambdaMu}, $P_{\mathrm{sc}}^{\mathrm{red}}=2\int_0^{T_p}\kappa|b|^2\,\dd t$. The real part of $1/\zeta$ gives the dispersive light shifts $\delta E_{g}=\hbar\OmP^2\D/[4(\D^2+\G^2/4)]$ and $\delta E_a=\hbar\OmS^2\D/[4(\D^2+\G^2/4)]$; the residual $\delta(t)$ in the main text is understood after compensation of their difference.

\section{Bare STIRAP: source equation, bound, and leading estimate}
\label{app:bare}

For $\Oma=\delta=0$, the reduced Hamiltonian projects onto the bright direction with complex energy $\hbar(\Lambda-\ii\kappa)$. Using $\ket{\dot D}=-\dot\theta\ket{B}$, $\ket{\dot B}=\dot\theta\ket{D}$ in the reduced Schr\"odinger equation gives the moving-basis equations
\begin{equation}
  \dot d=-\dot\theta\,b,
  \qquad
  \dot b=\dot\theta\,d-\mu b,
  \qquad
  \mu=\kappa+\ii\Lambda .
  \label{eq:bare-db}
\end{equation}
For $b(0)=0$, $b(t)=\int_0^t K(t,s)\dot\theta(s)d(s)\,\dd s$ with $K(t,s)=\exp[-\int_s^t\mu\,\dd u]$ and $|K|\le1$ since $\kappa\ge0$; hence $|b(t)|\le\int_0^t|\dot\theta|\,\dd s$. Integrating by parts using $\partial_s K=\mu K$ gives the exact identity
\begin{equation}
  b=\frac{\dot\theta d}{\mu}-\frac{\dot\theta(0)d(0)}{\mu(0)}K(t,0)-\int_0^t K\,\frac{\dd}{\dd s}\!\Bigl[\frac{\dot\theta d}{\mu}\Bigr]\dd s,
\end{equation}
whose leading local-response term is $b\simeq\dot\theta d/\mu\simeq\dot\theta/\mu$ (with $d\simeq1$), valid when $|\dot q|\ll|\mu q|$ for $q=\dot\theta d/\mu$. Evaluating the prefactor,
\begin{equation}
  |\mu|^2=\frac{\Omega^4}{16(\D^2+\G^2/4)},
  \qquad
  \frac{2\kappa}{|\mu|^2}=\frac{4\G}{\Omega^2},
\end{equation}
so that $P_{\mathrm{sc}}^{\mathrm{bare}}\simeq4\G\int_0^{T_p}\dot\theta^2/\Omega^2\,\dd t$, Eq.~\eqref{eq:Pbare}.

\section{Unified residual source and the counterdiabatic condition}
\label{app:source-full}

With the convention of Eq.~\eqref{eq:Ha}, $H_a=\ii\hbar\tfrac{\Oma}{2}(\ketbra{D}{B}-\ketbra{B}{D})$, so $\bra{B}H_a\ket{D}=-\ii\hbar\Oma/2$. The detuning term $H_\delta=\hbar\delta\ketbra{a}{a}$ contributes, via $\ket{a}=-\sin\theta\ket{D}+\cos\theta\ket{B}$, an off-diagonal $-\hbar\delta\sin\theta\cos\theta(\ketbra{D}{B}+\ketbra{B}{D})$ and a diagonal $\hbar\delta\cos^2\theta\ketbra{B}{B}$. Projecting the reduced Schr\"odinger equation onto $\ket{B}$ in the moving basis collects these with the geometric term and the bright energy to give Eq.~\eqref{eq:b-source} with the source Eq.~\eqref{eq:Fsource}. The real part $\dot\theta-\Oma/2$ is cancelled by Eq.~\eqref{eq:CDcond}; the imaginary part $\delta\sin\theta\cos\theta$ survives as the Doppler source. In the local-response regime, $b\simeq F/[\mu+\ii\delta\cos^2\theta]\simeq F/\mu$ for $|\delta|\ll|\mu|$, and $P_{\mathrm{sc}}^{\mathrm{primary}}\simeq4\G\int_0^{T_p}|F|^2/\Omega^2\,\dd t$, Eq.~\eqref{eq:Pprimary}. For completeness, the same residual source is derived in the full three-level model, without the local Raman elimination, in Appendix~\ref{app:exact}.

\section{Two-quadrature construction and equivalence to chirping}
\label{app:nulling}

The generalized coupling, Eq.~\eqref{eq:Hq}, has $y$-quadrature equal to the counterdiabatic term and $x$-quadrature $H_x=\hbar\tfrac{\Omega_x}{2}(\ketbra{D}{B}+\ketbra{B}{D})$. Since $\bra{B}H_x\ket{D}=\hbar\Omega_x/2$ is real, it adds an imaginary source $-\Omega_x/2$ to $F$, giving Eq.~\eqref{eq:Fq}. Choosing $\Omega_y=2\dot\theta$ and $\Omega_x=2\delta_c\sin\theta\cos\theta$ cancels both quadratures for $\delta=\delta_c$.

The cost of the construction, and the origin of its fragility, is visible in the bare $\{g,a\}$ basis. Using $\ketbra{D}{B}+\ketbra{B}{D}=\bigl(\begin{smallmatrix}\sin2\theta&\cos2\theta\\ \cos2\theta&-\sin2\theta\end{smallmatrix}\bigr)$, the $x$-quadrature is
\begin{equation}
  H_x=\frac{\hbar\delta_c\sin2\theta}{2}
  \begin{pmatrix}\sin2\theta & \cos2\theta\\[0.2em]\cos2\theta & -\sin2\theta\end{pmatrix},
  \label{eq:Hx-ga}
\end{equation}
which is not a pure $g$--$a$ coupling: it carries diagonal energy modulations $\pm\hbar\delta_c\sin^2 2\theta/2$ on $\ket{g}$ and $\ket{a}$. For the matched class this modulation, together with the full physical detuning $\delta_c$ retained in the response rate $\mu+\ii\delta_c\cos^2\theta$, is arranged to keep $b=0$. For a mismatched class $\delta=\delta_c+\varepsilon$ the residual source is $\ii\varepsilon\sin\theta\cos\theta$ --- identical to a chirp --- but the Hamiltonian still contains the fixed $\order(\delta_c)$ terms of Eq.~\eqref{eq:Hx-ga} and the detuning $\delta_c$. The chirp instead replaces $\delta\to\delta-\delta_c$, leaving only an $\order(\varepsilon)$ perturbation. At leading order $b\simeq F/\mu$ the two are indistinguishable [Eq.~\eqref{eq:Pnull}], but the local-response approximation requires $\delta_c\ll|\mu|$; once $\delta_c\sim|\mu|$ the fixed perturbation is no longer suppressed by the bright-state gap and the nulling departs from Eq.~\eqref{eq:Pnull}, as shown in Fig.~\ref{fig:doppler-chirp}. The single complex shortcut $\Omega_q\ee^{\ii\phi_q}=\Omega_x+\ii\Omega_y$ has magnitude $\Omega_q=2\sqrt{\dot\theta^2+\delta_c^2\sin^2\theta\cos^2\theta}$ and phase $\phi_q=\arctan[\dot\theta/(\delta_c\sin\theta\cos\theta)]$, so the construction requires both amplitude and phase control of the lower-state field.

\section{Optical reconstruction and auxiliary scattering cost}
\label{app:aux}

An auxiliary Raman pair $G_g,G_a$ through an excited state with $\Delta_R,\Gamma_R$ scatters, by the same elimination that produced Eq.~\eqref{eq:Psc-kappa}, with the functional
\begin{equation}
  P_{\mathrm{sc}}^{\mathrm{aux}}
  =\frac{\Gamma_R}{4(\Delta_R^2+\Gamma_R^2/4)}\int_0^{T_p}|G_gc_g+G_ac_a|^2\,\dd t .
  \label{eq:redscat-app}
\end{equation}
For a balanced scheme $G_g=G$, $G_a=-\ii G$ along the dark path $c_g=\cos\theta$, $c_a=-\sin\theta$, $|G_gc_g+G_ac_a|^2=G^2$, and the reconstruction condition $G^2=4\Delta_R\dot\theta$ gives $P_{\mathrm{aux}}=\tfrac{\Gamma_R\Delta_R}{\Delta_R^2+\Gamma_R^2/4}\int_0^{T_p}\dot\theta\,\dd t$. With $\int_0^{T_p}\dot\theta\,\dd t=\pi/2$ this is Eq.~\eqref{eq:Paux}. A power-limited maximum balanced scale $G_{\max}^2=4\Delta_R\dot\theta_{\max}$ gives $P_{\mathrm{aux,min}}\simeq2\pi\Gamma_R\dot\theta_{\max}/G_{\max}^2$.

\section{Mode-error budget, Doppler curvature, and contrast}
\label{app:budget}

The no-jump norm gives $N(T_p)=1-P_{\mathrm{sc}}$, and the conditional and unconditional mode fidelities are $F_{\mathrm{cond}}=|c_a(T_p)|^2/N(T_p)$ and $F_{\mathrm{mode}}=N(T_p)F_{\mathrm{cond}}\simeq F_{\mathrm{cond}}\ee^{-P_{\mathrm{sc}}}$. For $\epsilon_{\mathrm{coh}}=1-F_{\mathrm{cond}}\ll1$ and $P_{\mathrm{sc}}\ll1$, $1-F_{\mathrm{mode}}\simeq\epsilon_{\mathrm{coh}}+P_{\mathrm{sc}}$, with $P_{\mathrm{sc}}=P_{\mathrm{primary}}^{\mathrm{resid}}+P_{\mathrm{aux}}$. A finite Doppler width makes $F_{\mathrm{mode}}$ a function of the residual detuning $\delta_D$; to quadratic order $1-F_{\mathrm{mode}}(\delta_D)\simeq[1-F_{\mathrm{mode}}(0)]+A_D^{\mathrm{mode}}\delta_D^2$ with
\begin{align}
  A_D^{\mathrm{mode}}&=M_D+4\G J_D,\\
  M_D&=\Bigl|\int_0^{T_p}K(T_p,s)\sin\theta\cos\theta\,\dd s\Bigr|^2,
\end{align}
where the first term is coherent-endpoint curvature and the second is the scattering curvature from $J_D$ [Eq.~\eqref{eq:PscD}]. For a Gaussian ensemble of width $\sigma_D$, $\bar F_{\mathrm{mode}}=F_0/\sqrt{1+2A_D^{\mathrm{mode}}\sigma_D^2}\simeq F_0(1-A_D^{\mathrm{mode}}\sigma_D^2)$. Summing the additive single-pulse terms gives Eq.~\eqref{eq:mode-budget}; for $N_\pi$ independent pulses, $C/C_0\simeq\prod_j(1-\epsilon_{\mathrm{mode}}^{(j)})\simeq\exp[-\sum_j\epsilon_{\mathrm{mode}}^{(j)}]$, and a target contrast imposes $\sum_j\epsilon_{\mathrm{mode}}^{(j)}\le-\ln C_{\mathrm{target}}$.

\section{Gaussian mixing-rate schedule}
\label{app:gaussian}

All figures use $\dot\theta(t)=A\exp[-(t-T_p/2)^2/2\sigma^2]$ for $0\le t\le T_p$, with $A$ fixed by $\int_0^{T_p}\dot\theta\,\dd t=\pi/2$,
\begin{equation}
  A=\frac{\pi/2}{\sqrt{2\pi}\,\sigma\,\mathrm{erf}[T_p/(2\sqrt2\sigma)]} .
\end{equation}
The squared-rate integral entering Eq.~\eqref{eq:Pbare} is $I_\theta=\int_0^{T_p}\dot\theta^2\,\dd t=\tfrac{\pi^2}{8\sqrt\pi\sigma}\,\mathrm{erf}[T_p/(2\sigma)]/\mathrm{erf}^2[T_p/(2\sqrt2\sigma)]$, so for constant $\Omega=\Omega_0$, $P_{\mathrm{sc}}^{\mathrm{bare}}\simeq(4\G/\Omega_0^2)I_\theta$.

\section{Full three-level no-jump benchmark equations}
\label{app:numerics}

The benchmark integrates the full no-jump equations in the convention of Eq.~\eqref{eq:H3}:
\begin{align}
  \dot c_g &= -\ii\tfrac{\OmP}{2}c_e+\tfrac{\Oma}{2}c_a,\\
  \dot c_e &= -\ii\tfrac{\OmP}{2}c_g+\ii\D c_e-\tfrac{\G}{2}c_e-\ii\tfrac{\OmS}{2}c_a,\\
  \dot c_a &= -\tfrac{\Oma}{2}c_g-\ii\tfrac{\OmS}{2}c_e-\ii\delta c_a,\\
  \dot P_{\mathrm{sc}}^{(3)} &= \G|c_e|^2,\qquad P_{\mathrm{sc}}^{(3)}(0)=0 .
\end{align}
These reproduce the exact identity Eq.~\eqref{eq:Psc-exact} without invoking the local Raman elimination or the dark--bright reduction. With this convention, the counterdiabatic condition $\Oma=2\dot\theta$ cancels the source; the chirp protocol replaces $\delta\to\delta-\delta_c$, and the two-quadrature nulling adds the lower-state block of Eq.~\eqref{eq:Hx-ga} together with $\Omega_y=2\dot\theta$. Figures~\ref{fig:bare-hierarchy}--\ref{fig:doppler-chirp} were generated from these equations using an adaptive Runge--Kutta integrator with absolute and relative tolerances at the $10^{-9}$--$10^{-12}$ level.

\section{Exact three-level structure without adiabatic elimination}
\label{app:exact}

Appendices~\ref{app:local}--\ref{app:source-full} establish the source identity and the
residual source $F$ after the local Raman elimination of $\ket{e}$. Here we show that the
same identity $S=\Omega b$, the same source $F$, and the counterdiabatic cancellation all
hold \emph{exactly} in the full no-jump dynamics, with $\ket{e}$ retained; the only new
object is the exact moving-frame generator.

We rotate the full no-jump Hamiltonian
$H_{\mathrm{nh}}=H_3-\tfrac{\ii\hbar\G}{2}\ketbra{e}{e}$ [Eq.~\eqref{eq:H3}] into the
instantaneous dark--bright--excited basis $\{\ket{D},\ket{B},\ket{e}\}$
[Eq.~\eqref{eq:DB}], holding $\ket{e}$ fixed, through
$\mathcal{H}=U^{\dagger}H_{\mathrm{nh}}U-\ii\hbar\,U^{\dagger}\dot U$ with matrix elements
$\mathcal{H}_{nm}=\bra{n}H_{\mathrm{nh}}\ket{m}-\ii\hbar\braket{n}{\dot m}$. The rotation
$U(t)=\ketbra{D}{g}+\ketbra{B}{a}+\ketbra{e}{e}
=\exp\!\big[\theta(\ketbra{g}{a}-\ketbra{a}{g})\big]$ carries the lower states into the
instantaneous dark--bright pair while fixing $\ket{e}$, and generates the nonadiabatic
connection $U^{\dagger}\dot U=\dot\theta(\ketbra{D}{B}-\ketbra{B}{D})$.

The four contributions to $\mathcal{H}$ are then explicit. The optical couplings give
$\bra{e}H_3\ket{D}=\tfrac{\hbar}{2}(\OmP\cos\theta-\OmS\sin\theta)=0$ and
$\bra{e}H_3\ket{B}=\tfrac{\hbar}{2}(\OmP\sin\theta+\OmS\cos\theta)=\hbar\Omega/2$, the
exact optical darkness of $\ket{D}$. The connection contributes $\mp\ii\hbar\dot\theta$
to the $(D,B)$ and $(B,D)$ entries via $\braket{D}{\dot B}=\dot\theta=-\braket{B}{\dot D}$.
The shortcut [Eq.~\eqref{eq:Ha}] gives $\bra{B}H_a\ket{D}=-\ii\hbar\Oma/2$, using
$\ketbra{g}{a}-\ketbra{a}{g}=\ketbra{D}{B}-\ketbra{B}{D}$. The detuning
$H_\delta=\hbar\delta\ketbra{a}{a}$, with $\ket{a}=-\sin\theta\ket{D}+\cos\theta\ket{B}$,
contributes the diagonal terms $\hbar\delta\sin^2\theta$ and $\hbar\delta\cos^2\theta$
together with the off-diagonal $-\hbar\delta\sin\theta\cos\theta$. Collecting these, the
dark--bright entry is
$\mathcal{H}_{BD}=\ii\hbar(\dot\theta-\Oma/2)-\hbar\delta\sin\theta\cos\theta=\ii\hbar F$,
with $F$ exactly the source of Eq.~\eqref{eq:Fsource} and
$\mathcal{H}_{DB}=\mathcal{H}_{BD}^{*}=-\ii\hbar F^{*}$, so that
\begin{equation}
  \frac{\mathcal{H}}{\hbar}=
  \begin{pmatrix}
    \delta\sin^2\theta & -\ii F^{*} & 0\\[3pt]
    \ii F & \delta\cos^2\theta & \Omega/2\\[3pt]
    0 & \Omega/2 & -\D-\tfrac{\ii}{2}\G
  \end{pmatrix}_{(D,B,e)} .
  \label{eq:Hcal-exact}
\end{equation}
The dark--bright block is Hermitian, and the only non-Hermitian element is the
$-\tfrac{\ii}{2}\G$ on $\ket{e}$. No elimination has been made: the excited amplitude is
retained through the $\Omega/2$ couplings, which is the sole difference from the reduced
generator $H_{\mathrm{red}}$ of Appendix~\ref{app:source}.

The equations of motion follow directly. Reading
$\ii\dot{\mathbf a}=(\mathcal{H}/\hbar)\,\mathbf a$ off Eq.~\eqref{eq:Hcal-exact} with
$\mathbf a=(d,b,c_e)^{\mathsf T}$ and $\zeta=\D+\tfrac{\ii}{2}\G$,
\begin{equation}
  \begin{aligned}
    \ii\dot d &= \delta\sin^2\theta\,d-\ii F^{*}b,\\
    \ii\dot b &= \ii F\,d+\delta\cos^2\theta\,b+\tfrac{\Omega}{2}c_e,\\
    \ii\dot c_e &= \tfrac{\Omega}{2}\,b-\zeta\,c_e,
  \end{aligned}
  \label{eq:eom-exact}
\end{equation}
which are Eqs.~\eqref{eq:cg}--\eqref{eq:ca} in the moving basis. The excited state is
sourced by $b$ alone, the exact content of the identity $S=\Omega b$
[Eq.~\eqref{eq:Sbright}]; this holds here without approximation because the identity is a
statement about the basis, not a consequence of elimination.

The third of Eqs.~\eqref{eq:eom-exact} integrates exactly to the memory kernel of
Appendix~\ref{app:local} [Eq.~\eqref{eq:ce-memory} with $c_e(0)=0$ and $S=\Omega b$],
$c_e(t)=-\tfrac{\ii}{2}\int_0^t\ee^{\ii\zeta(t-s)}\Omega(s)b(s)\,\dd s$. Substituting it
into the bright equation eliminates $c_e$ in favor of the bright history and gives the
exact, non-Markovian, closed pair
\begin{equation}
  \begin{aligned}
    \ii\dot b &= \ii F\,d+\delta\cos^2\theta\,b -\frac{\ii\,\Omega(t)}{4}\int_0^t\ee^{\ii\zeta(t-s)}\Omega(s)\,b(s)\,\dd s,\\
    \ii\dot d &= \delta\sin^2\theta\,d-\ii F^{*}b.
  \end{aligned}
  \label{eq:b-exact-int}
\end{equation}
The bright amplitude is driven by $F$ and damped by the retarded self-energy from its own
coupling to $\ket{e}$; the local form $b\simeq F/\mu$ and the reduced functional
[Eqs.~\eqref{eq:b-source} and \eqref{eq:Psc-kappa}] follow only after the Markov
reduction of Appendix~\ref{app:local}.

The intrinsic decay of the bright amplitude is set by the same coupling. On two-photon
resonance and for locally constant $\Omega$, the $\{B,e\}$ sector of
Eq.~\eqref{eq:Hcal-exact} has eigenvalues
$E_\pm=\tfrac12[-\D-\tfrac{\ii}{2}\G\pm\sqrt{W}]$ with
$W=(\D+\tfrac{\ii}{2}\G)^2+\Omega^2$, so the bright-like mode decays at the exact
power-broadened rate
\begin{equation}
  \kappa_\star=-\operatorname{Im}E_+
  =\frac{\G}{4}-\frac12\sqrt{\frac{|W|-\operatorname{Re}W}{2}} ,
  \label{eq:kstar}
\end{equation}
which reduces to $\kappa$ [Eq.~\eqref{eq:kappaLambdaMu}] for $\Omega\ll|\zeta|$ and
saturates at $\G/4$ for strong driving. Equation~\eqref{eq:kstar} is the exact linewidth
at which $b$ leaks into $\ket{e}$ and fixes the validity of the rate $\kappa$ used in
Eq.~\eqref{eq:Psc-kappa}: the reduction degrades once $\Omega$ approaches saturation,
where $\kappa\propto\Omega^2$ grows without bound while $\kappa_\star\to\G/4$.

Finally, the cancellation is exact. At the counterdiabatic point Eq.~\eqref{eq:CDcond}
with $\delta=0$ the source vanishes, $F\equiv0$, and $\mathcal{H}_{DD}=0$, so
Eq.~\eqref{eq:Hcal-exact} block-diagonalizes into a decoupled dark amplitude and a
homogeneous $\{B,e\}$ pair. For an initially dark state ($d(0)=1$, $b(0)=c_e(0)=0$) the
unique solution is $b(t)=c_e(t)=0$ for all $t$, so Eq.~\eqref{eq:Psc-exact} gives
$\Psc=0$ identically, for arbitrary $\D$, $\Omega$, $\G$, and $T_p$. This is the exact,
elimination-free counterpart of Eq.~\eqref{eq:Pprimary-zero}: the cancellation follows
structurally from the optical darkness $\bra{e}H_3\ket{D}=0$ and the lower-manifold
action of $H_a$, with no expansion in $1/\D$ or $\G/\Omega$ of the kind underlying
Appendices~\ref{app:local}--\ref{app:source-full}.

\bibliography{bibliography}

\end{document}